\documentclass[dvips]{acta}
\usepackage{supertabular,lscape,epsfig}
\usepackage{amssymb}
\usepackage{amsmath}
\usepackage{wasysym}
\DeclareSymbolFont{ppa}{OT1}{ppl}{m}{it}
\DeclareMathSymbol{\vv}{\mathalpha}{ppa}{'166}

\newfont{\hb}{rphvb at 10pt}
\newfont{\hbo}{rphvbo at 10pt}
\newfont{\bitt}{rptmbi at 12pt}
\newfont{\bits}{rptmbi at 11pt}

\SetPages{1}{1}

\SetVol{61}{2011}

\begin{document}

\newcommand{\TabApp}[2]{\begin{center}\parbox[t]{#1}{\centerline{
  {\bf Appendix}}
  \vskip2mm
  \centerline{\small {\spaceskip 2pt plus 1pt minus 1pt T a b l e}
  \refstepcounter{table}\thetable}
  \vskip2mm
  \centerline{\footnotesize #2}}
  \vskip3mm
\end{center}}

\newcommand{\TabCapp}[2]{\begin{center}\parbox[t]{#1}{\centerline{
  \small {\spaceskip 2pt plus 1pt minus 1pt T a b l e}
  \refstepcounter{table}\thetable}
  \vskip2mm
  \centerline{\footnotesize #2}}
  \vskip3mm
\end{center}}

\newcommand{\TTabCap}[3]{\begin{center}\parbox[t]{#1}{\centerline{
  \small {\spaceskip 2pt plus 1pt minus 1pt T a b l e}
  \refstepcounter{table}\thetable}
  \vskip2mm
  \centerline{\footnotesize #2}
  \centerline{\footnotesize #3}}
  \vskip1mm
\end{center}}

\newcommand{\MakeTableApp}[4]{\begin{table}[p]\TabApp{#2}{#3}
  \begin{center} \TableFont \begin{tabular}{#1} #4 
  \end{tabular}\end{center}\end{table}}

\newcommand{\MakeTableSepp}[4]{\begin{table}[p]\TabCapp{#2}{#3}
  \begin{center} \TableFont \begin{tabular}{#1} #4 
  \end{tabular}\end{center}\end{table}}

\newcommand{\MakeTableee}[4]{\begin{table}[htb]\TabCapp{#2}{#3}
  \begin{center} \TableFont \begin{tabular}{#1} #4
  \end{tabular}\end{center}\end{table}}

\newcommand{\MakeTablee}[5]{\begin{table}[htb]\TTabCap{#2}{#3}{#4}
  \begin{center} \TableFont \begin{tabular}{#1} #5 
  \end{tabular}\end{center}\end{table}}

\newfont{\bb}{ptmbi8t at 12pt}
\newfont{\bbb}{cmbxti10}
\newfont{\bbbb}{cmbxti10 at 9pt}
\newcommand{\uprule}{\rule{0pt}{2.5ex}}
\newcommand{\douprule}{\rule[-2ex]{0pt}{4.5ex}}
\newcommand{\dorule}{\rule[-2ex]{0pt}{2ex}}
\def\thefootnote{\fnsymbol{footnote}}

\hyphenation{OSARGs}

\begin{Titlepage}

\Title{The Optical Gravitational Lensing Experiment.\\
The OGLE-III Catalog of Variable Stars.\\
XIII. Long-Period Variables in the Small Magellanic Cloud\footnote{Based on
observations obtained with the 1.3-m Warsaw telescope at the Las Campanas
Observatory of the Carnegie Institution for Science.}}
\Author{I.~~S~o~s~z~y~ñ~s~k~i$^1$,~~
A.~~U~d~a~l~s~k~i$^1$,~~
M.\,K.~~S~z~y~m~a~ñ~s~k~i$^1$,~~
M.~~K~u~b~i~a~k$^1$,\\
G.~~P~i~e~t~r~z~y~ñ~s~k~i$^{1,2}$,~~
£.~~W~y~r~z~y~k~o~w~s~k~i$^{1,3}$,~~
K.~~U~l~a~c~z~y~k$^1$,~~
R.~~P~o~l~e~s~k~i$^1$,\\
S.~~K~o~z~³~o~w~s~k~i$^1$
~~and~~ P.~~P~i~e~t~r~u~k~o~w~i~c~z$^1$}
{$^1$Warsaw University Observatory, Al.~Ujazdowskie~4, 00-478~Warszawa, Poland\\
e-mail:
(soszynsk,udalski,msz,mk,pietrzyn,kulaczyk,rpoleski,simkoz,pietruk)@astrouw.edu.pl\\
$^2$ Universidad de Concepción, Departamento de Astronomia, Casilla 160--C, Concepción, Chile\\
$^3$ Institute of Astronomy, University of Cambridge, Madingley Road, Cambridge CB3 0HA, UK\\
e-mail: wyrzykow@ast.cam.ac.uk}
\Received{September 6, 2011}
\vspace*{-6pt}
\end{Titlepage}
\vspace*{-9pt}
\Abstract{The thirteenth part of the OGLE-III Catalog of Variable Stars
(OIII-CVS) contains $19\;384$ long-period variables (LPVs) detected in the
Small Magellanic Cloud. The sample is composed of 352 Mira stars, 2222
semiregular variables (SRVs) and $16\;810$ OGLE Small Amplitude Red Giants
(OSARGs). Sources are divided into oxygen-rich and carbon-rich stars. The
catalog includes time-series {\it VI} photometry obtained between 1997 and
2009.

Methods used to select and classify variable stars are described. We show
some statistical properties of the sample, and compare it with LPVs in the
Large Magellanic Cloud. Additionally, we present objects of particular
interest, \eg a SRV with outbursts, and a Mira star with the longest known
pulsation period $P=1860$~days.}
{Stars: AGB and post-AGB -- Stars: late-type -- Stars: oscillations --
Magellanic Clouds}

\Section{Introduction}

When low to intermediate mass stars climb the red giant branch (RGB) or the
asymptotic giant branch (AGB), they exhibit various forms of intrinsic
variability. In the first phase of the evolution as red giants the
stochastically excited solar-like oscillations are excited. The amplitudes
of these variations increase with the brightness of stars and near the tip
of the RGB (for both, RGB and AGB stars) solar-like oscillations can be
observed from the ground as OGLE Small Amplitude Red Giants (OSARGs,
Dziembowski and Soszyñski 2010). When stars are close to end of their
evolution on the AGB, they start pulsations as semiregular variables (SRVs)
or Miras. All these kinds of such objects are called long-period variables
(LPVs). LPVs are common in various stellar systems, they are among the
brightest stars and obey multiple period-luminosity (PL) relations, which
makes them potentially important as distance indicators. Variability of red
giants also provides information about the physical structure of these
stars and probe their mass loss rates. Evolutionary advanced AGB stars are
very effective contributors of re-cycled material to the interstellar
medium.

The Large (LMC) and Small Magellanic Clouds (SMC) offer an opportunity to
compare huge samples of red giant stars in the environments that exhibit
different metallicities. In the fourth paper of this series (Soszyñski
\etal 2009, hereafter Paper~I) we described the catalog of nearly $92\;000$
LPVs in the LMC compiled from the observations collected during the third
phase of the Optical Gravitational Lensing Experiment (OGLE-III). Here we
present the OGLE-III catalog of $19\;384$ LPVs in the SMC -- the largest
such catalog published to date.

First several variable red giants in the SMC were identified as a
by-product of the Harvard survey for Cepheids in this galaxy (Shapley \etal
1925, Hoffleit 1935, McKibben Nail 1942). In those days these objects were
thought to be Galactic foreground LPVs. Only Shapley and McKibben Nail
(1951) showed that the concentration of LPVs toward the SMC center rules
out the possibility that all these objects lay in the foreground. They
provided a list of 42 Miras, semiregular and irregular variables in the
SMC. Then, several LPVs in the SMC were found during the surveys of
Dartayet and Landi Dessy (1952) and Thackeray (1958). The lists of 24
long-period and 61 irregular variables were provided in the catalog of
variable stars in the SMC by Payne-Gaposchkin and Gaposchkin (1966). A
survey devoted solely to the search for variable red giants in the
Magellanic Clouds were initiated by Lloyd Evans (1971). As a result several
dozen new LPVs in the SMC were discovered and analyzed (Lloyd Evans 1978ab,
Lloyd Evans \etal 1988).

Long-term observations from microlensing sky surveys, especially MACHO and
OGLE, provided unprecedented resources for variable red giant
research. LPVs in the SMC were studied by Cioni \etal (2003), Groenewegen
(2004), Ita \etal (2004ab), Kiss and Bedding (2004), Schultheis \etal
(2004), Soszyñski \etal (2004, 2007), Raimondo \etal (2005), Lah \etal
(2005), usually in comparison with the larger sample of LPVs in the LMC. No
catalog of LPVs in the SMC was published so far on the basis of photometric
data provided by the microlensing projects.

In this paper we describe the thirteenth part of the OGLE-III Catalog of
Variable Stars. The previous parts of the Catalog comprise over $173\;000$
variables stars of various types in the Magellanic Clouds and Galactic
bulge. In this study, like in Paper~I, we divide LPVs into three groups:
OSARGs (Wray \etal 2004), SRVs and Miras. Soszyñski \etal (2004) showed
that OSARGs and SRVs constitute a separate classes of LPVs, with different
series of PL relations and probably with different pulsation mechanism.

\vspace{5mm}
\Section{Observational Data}

The OGLE project provides a database of {\it I}- and {\it V}-band
time-series photometry of about 6 million stars in the SMC. The photometry
in the central parts of the SMC was collected over a period of nearly 13
years (OGLE-II + OGLE-III), from January 1997 to May 2009. The outer parts
of the SMC were monitored over a shorter time span -- from June
2001. Observations were obtained with the 1.3-m Warsaw telescope at Las
Campanas Observatory (operated by the Carnegie Institution for Science) in
Chile. During the OGLE-III survey the telescope was equipped with the
eight-chip CCD mosaic camera of the total resolution $8192\times8192$
pixels and the field of view of about $35'\times35\zdot\arcm5$. Details of
the instrumentation setup can be found in Udalski (2003).

The OGLE-III survey in the SMC covered an area of about 14 square degrees
(41 fields). Typically about 700 observing points were collected in the
Cousins {\it I} photometric band, and 50-70 measurements in the Johnson
{\it V} band. For stars for which the OGLE-II photometry is available the
number of observations is about 1000 and 100 in the {\it I} and {\it V}
bands, respectively. When a star was located in the overlapping part of
adjacent fields, the number of points may be larger, because we merged
observations from both fields.

The {\it VI} photometry was obtained with the standard OGLE data reduction
pipeline (Udalski et al. 2008) based on the Difference Image Analysis (DIA,
Alard and Lupton 1998, Wo¼niak 2000). For 22 bright stars, which saturate
in the DIA reference images, we provide the {\sc DoPhot} {\it I}-band
photometry (Schechter \etal 1993). These stars are flagged in the remarks
of the catalog.

We matched OGLE stars against the near-infrared 2MASS All-Sky Catalog of
Point Sources (Cutri \etal 2003) and the IRSF Magellanic Clouds Point
Source Catalog (Kato \etal 2007) using a search radius of 1 arcsec. In this
analysis we use reddening-free Wesenheit magnitudes, defined as:
$$W_{JK}=K_s-0.686(J-K_s)$$
$$W_I=I-1.55(V-I)$$ In these formulas $J$ and $K_s$ are single-epoch
near-infrared measurements taken from the 2MASS or IRSF catalogs, while
$W_I$ was derived independently for each pair of measurements in $V$ and
$I$ bands, and the mean value of $W_I$ was derived for a given star.

\Section{Selection and Classification of LPVs}

The selection of LPVs in the SMC began with the massive period search
performed for all stars monitored by the OGLE-III project in this
galaxy. We search a frequency space of 0 to 0.5~day$^{-1}$ with a frequency
spacing of 0.000001~ day$^{-1}$ using the {\sc Fnpeaks} code (Ko³aczkowski,
private communication). Fifteen periods with the corresponding amplitudes
were recorded for each star with an iterative procedure of determining
periods and prewhitening the light curves with these periods.

LPVs in the SMC were selected in the same way as in the LMC (Paper
I). Large- and medium-amplitude light curves (SRVs and Miras) were visually
inspected and the classification was done based on the light curve
morphology, position of a star in the PL, color-magnitude and
period-amplitude diagrams. Like in the catalog of LPVs in the LMC, we
inspected light curves of very faint objects ($I\approx20$~mag) and
detected a number of Miras and SRVs surrounded by circumstellar
matter. Most of the small-amplitude LPVs (OSARGs) were identified
automatically using the method described by Soszyñski \etal (2007). The
method utilizes the positions of OSARGs in the reddening-free PL diagrams
and the characteristic period-ratios exhibited by these multi-periodic
objects.

In some stars the automatically determined primary periods were found to be
spurious and they were replaced by one of the secondary periods, usually
corresponding to one of the PL sequences discovered by Wood \etal
(1999). Note that only the primary period for each star has been
verified. The second and the third periods, also given in this catalog, may
be false. For detailed analyses of the periodicity we recommend use
original photometry provided with the catalog. The red giants in eclipsing
and ellipsoidal binary systems, which form sequence~E in the PL plane, were
included in this catalog only if they simultaneously showed pulsations as
OSARGs or SRVs. Main periods of these stars listed in our catalog usually
correspond to half of the orbital periods.

\begin{figure}[htb]
\centerline{\includegraphics[width=11.3cm, bb=65 270 555 745]{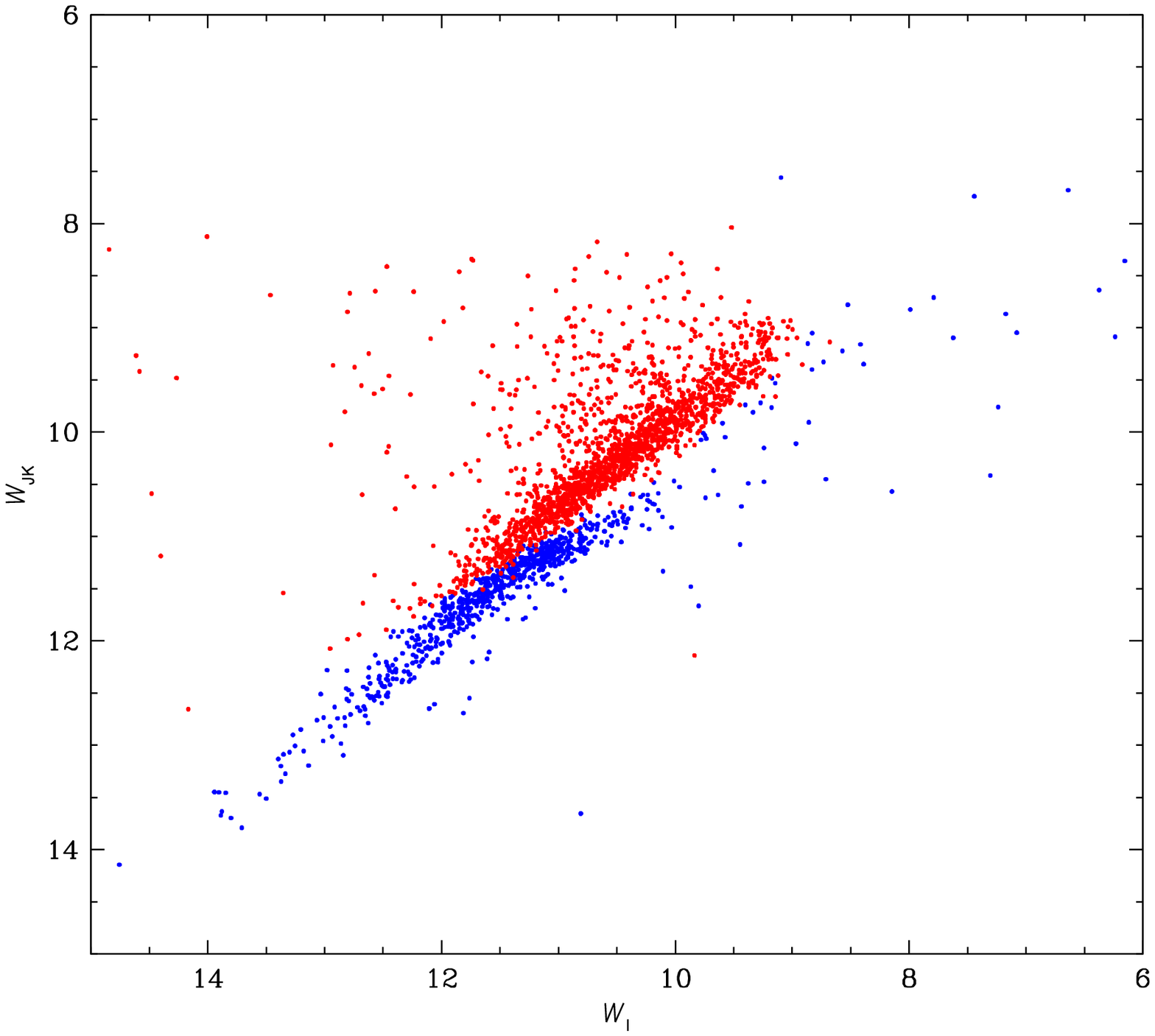}}
\FigCap{$W_I$--$W_{JK}$ diagram for Miras and SRVs in the SMC. Blue points
show O-rich, while red points represent C-rich stars.}
\end{figure}

With the aim of excluding foreground Galactic objects from our catalog, we
checked the astrometric measurements of the initially selected sources, and
removed stars with detectable proper motions. Only in the field SMC140,
centered on the Galactic globular cluster 47~Tuc, SMC stars showed proper
motions relative to the cluster members (of course, we observed the motion
of 47~Tuc, but the coordinate grid was established on the cluster stars,
which dominate in the field SMC140). LPVs that are members of 47~Tuc are
saturated in the OGLE data, so they are not included in our catalog.

\begin{figure}[htb]
\centerline{\includegraphics[width=12cm, bb=55 100 555 750]{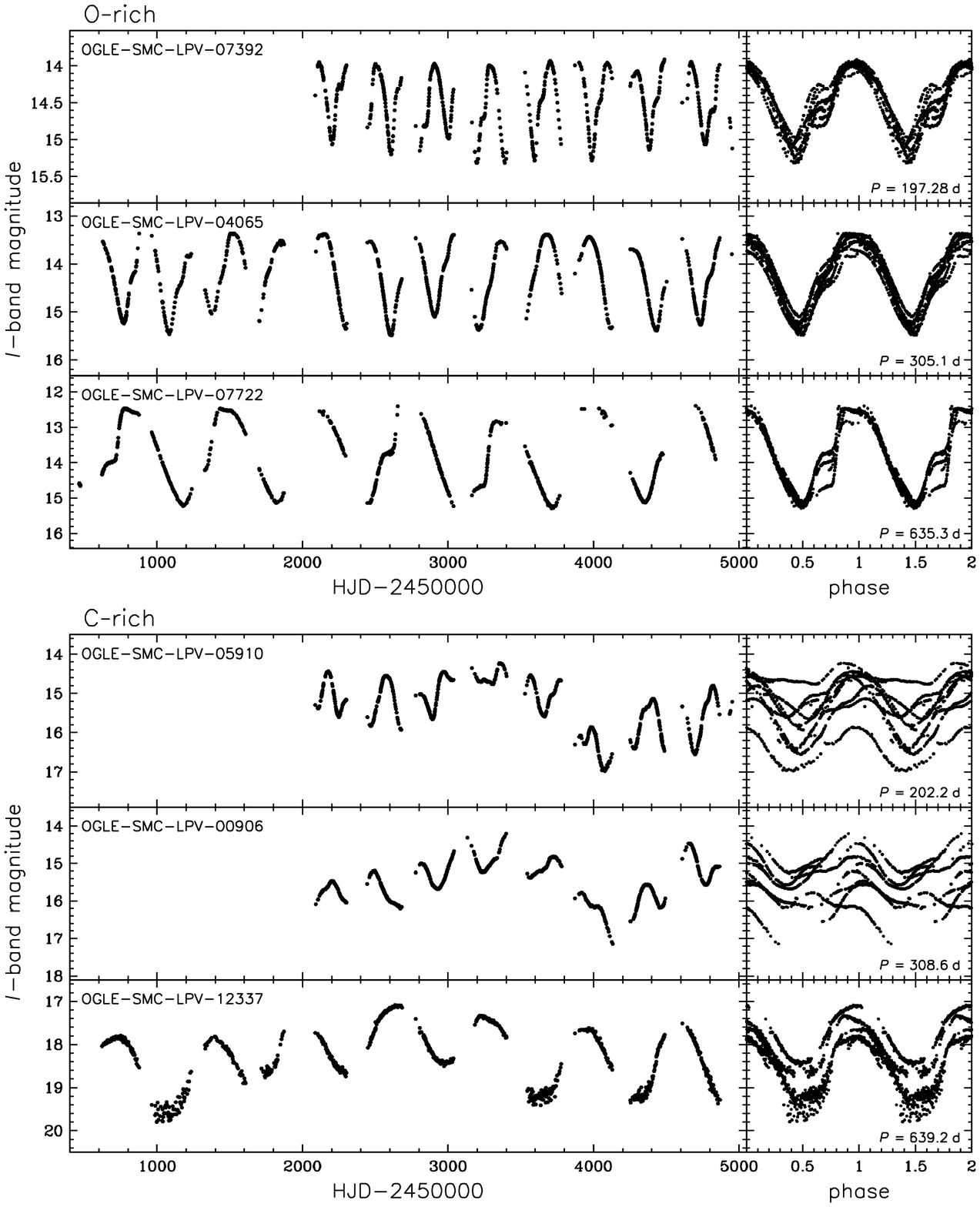}}
\FigCap{OGLE-II (if available) and OGLE-III light curves of three
typical O-rich Miras (three {\it upper panels}) and three C-rich Miras
(three {\it lower panels}). Note different light curves of both spectral
classes. {\it Left panels} show unfolded light curves, and {\it right
panels} show the same light curves folded with the primary periods.}
\end{figure}

LPVs were divided into carbon- (C-) and oxygen-rich (O-rich) stars using
several methods. The primary criterion was the position of the star in the
$W_I$--$W_{JK}$ diagram (Fig.~1). The same method was used to distinguish
between O-rich and C-rich giants in the LMC (Paper~I), however the
extension of the SMC along the line of sight is larger than the LMC, which
forced us to look also for other solutions. As the secondary tests we used
position of SRVs and Miras in the period--$W_I$ diagram (Soszyñski \etal
2005), $(J-K_s)$ colors, and the morphology of light curves. Fig.~2 shows
examples of typical light curves of O-rich and C-rich Miras. It is evident
that C-rich LPVs exhibit significantly larger changes of the mean
luminosity, which is likely connected with episodes of intensive mass
loss. For stars included in the {\sc Simbad} astronomical
database\footnote{http://simbad.u-strasbg.fr/simbad/} we compared our
photometric solution with the spectroscopic determinations of the stellar
types, and obtained a good agreement of both results.

\begin{figure}[htb]
\centerline{\includegraphics[width=11.7cm, bb=55 370 555 745]{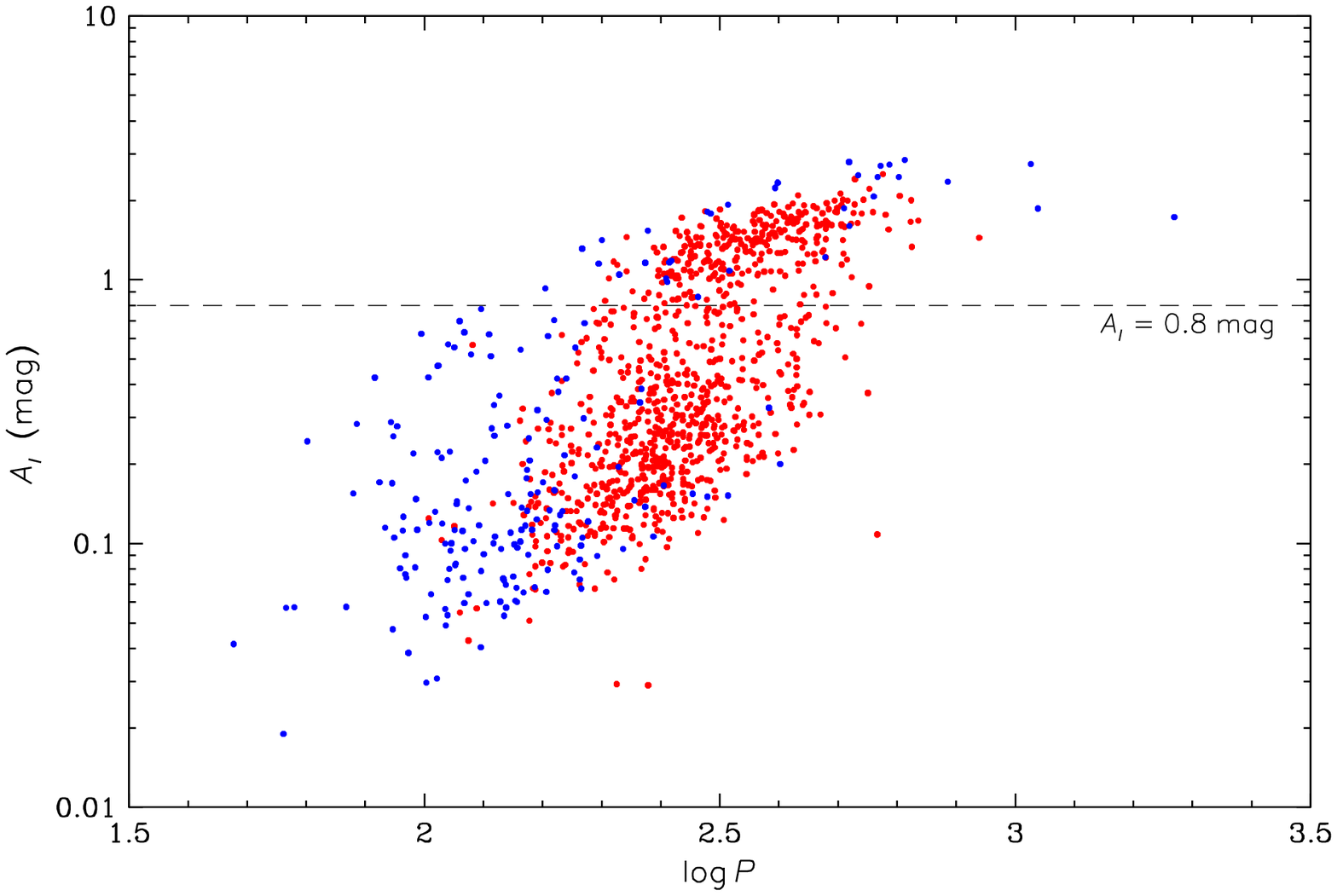}}
\FigCap{Period--amplitude diagram for the sequence C stars (Miras and
SRVs). Blue points show O-rich, while red points represent C-rich
stars. Black dashed line mark the limiting amplitude ($A_I=0.8$~mag) used
to separate Miras and SRVs in this paper.}
\end{figure}

To derive the amplitudes of the periodic variations we applied an iterative
procedure of detrending light curves from slow irregular variations, common
in C-rich variables. The period--amplitude diagram (Fig.~3) confirms a
conclusion presented in Paper~I that for C-rich stars there is a natural
boundary between Miras and SRVs. In this work we use the same definition of
Mira stars as in the catalog of LPVs in the LMC: Miras are pulsating stars
which lie in the PL sequence C and have {\it I}-band peak-to-peak
amplitudes larger than 0.8~mag. It is important to note that some stars
with periods in sequence D satisfy the amplitude criterion, however their
light curves have very different shapes than Mira stars. Sequence D is
populated by still unexplained long secondary periods observed in a large
fraction of SRVs and OSARGs.

In contrast to the catalog of LPVs in the LMC (Paper~I), we have not
divided our sample into RGB and AGB stars. Obviously, the OSARGs brighter
than the tip of the RGB ($K_s\approx12.7$~mag), and all SRVs and Miras are
stars evolving along the AGB. The OSARG variables below the tip of the RGB
are a mixture of RGB and AGB stars. Our method of distinguishing between
RGB and AGB OSARGs described by Soszyñski \etal (2004) worked well (at
least in the statistical sense) for the LMC variables, where the PL
relations are relatively well separated (however note the discussion about
this method by Tabur \etal 2010). The significant extension of the SMC
along the line of sight blurs the PL pattern and limits the application of
our method.

\vspace{5mm}
\Section{Catalog of LPVs in the SMC}

The catalog of LPVs in the SMC contains $19\;384$ objects, of which
$16\;810$ have been classified as OSARGs, 2222 as SRVs and 352 as
Miras. The proportion of different types of LPVs is more or less the same
as in the LMC (Paper~I). Both galaxies significantly differ in the relative
number of O-rich and C-rich stars. In the SMC about 30\% of SRVs and Miras
are O-rich stars and 70\% are C-rich giants. In the LMC SRVs and Miras
divide roughly equally to the O-rich and C-rich stars. Among the SMC OSARG
variables only a few percent of objects have been categorized as C-rich
giants.

\begin{figure}[htb]
\centerline{\includegraphics[width=12.7cm]{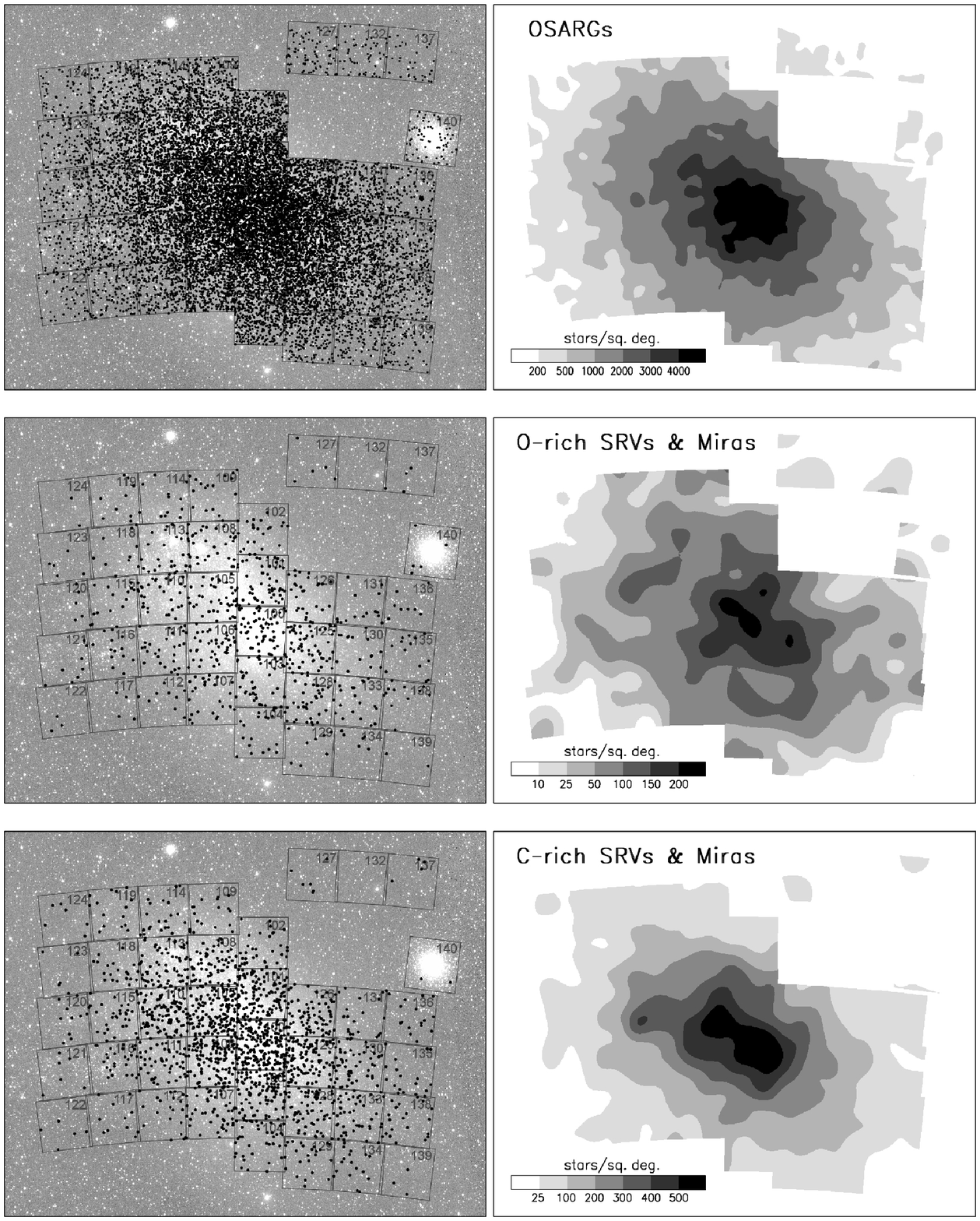}}
\vspace*{2mm}
\FigCap{Spatial distribution of LPVs in the SMC. {\it Left panels} 
show positions of stars overplotted on the SMC image originated from the
ASAS sky survey (Pojmañski 1997). Right panels present surface density maps
obtained by the smoothing of the distributions with the Gaussian
filter. {\it Upper panels} show OSARG variables, {\it middle panels} --
O-rich SRVs and Miras and {\it bottom panels} -- C-rich SRVs and Miras.}
\end{figure}

The catalog data are accessible through the anonymous FTP site or via the
web interface:
\begin{center}
{\it ftp://ftp.astrouw.edu.pl/ogle/ogle3/OIII-CVS/smc/lpv/}\\ {\it
http://ogle.astrouw.edu.pl/} \\
\end{center}

The FTP site is organized as follows. The file named {\sf ident.dat}
contains the full list of LPVs with their classification and identification
in other databases. The stars are organized in order of increasing right
ascension and designated with symbols OGLE-SMC-LPV-NNNNN, where NNNNN is a
five-digit consecutive number. The {\sf ident.dat} file contains the
following columns: the object designation, the OGLE-III field and the
internal database number of the object, classification (Mira, SRV, OSARG),
spectral type (O-rich, C-rich), equinox J2000.0 right ascension and
declination, cross-identifications with the OGLE-II database (Szymañski
2005), the MACHO photometric database, and the extragalactic part of the
GCVS (Artyukhina et al. 1995). Files {\sf OSARGs.dat}, {\sf SRVs.dat}, and
{\sf Miras.dat} contain observational parameters of the relevant types of
LPVs: mean magnitudes in the {\it I} and {\it V} bands, and three dominant
periods with peak-to-peak {\it I}-band amplitudes for each star.

The files containing time-series photometry in the {\it I} and {\it V}
bands are stored in the directory {\sf phot}. Finding charts  are in the
directory {\sf fcharts}. These are $60\arcs\times60\arcs$ subframes of the
{\it I}-band DIA reference images, oriented with North to the top and East
to the left. Additional information on some objects (long secondary
periods, eclipsing or ellipsoidal variability, uncertain classification,
etc.) are given in the file {\sf remarks.txt}.

Fig.~4 presents the spatial distribution of LSPs in the SMC, separately for
OSARGs, O-rich and C-rich SRV and Miras. The left panels show position of
individual objects superimposed on the SMC image obtained from the ASAS
project (Pojmañski 1997), while the right panels display the same
distribution smoothed with the Gaussian filter. Due to the low statistics
it is difficult to clearly identify the differences between the
distributions of O- and C-rich variables, as it was possible in the LMC
(Paper~I). However, it seems that the distribution of the C-rich SRVs and
Miras in the sky is more elongated then the O-rich LPVs, as it was found in
the LMC.

We judge the level of completeness of this catalog to be the same as in the
catalog of LPVs in the LMC (Paper~I). The comparison with the GCVS and the
list of LPVs published by Raimondo \etal (2005) shows that we have found
virtually all large and medium-amplitude variables in the OGLE-III fields,
with the exception of several brightest stars, which are saturated in our
database. The completeness decreases for OSARG variables with the smallest
amplitudes ($A_I\apprle0.005$~mag), \ie the faintest ones. The solar-like
oscillations with amplitudes much below the OGLE detection limits were
observed in red giants by CoRoT and Kepler space telescopes (De Ridder
\etal 2009, Gilliland, R.L., \etal 2010). Tabur \etal (2010) demonstrated a
continuous transitions between these very low-amplitude oscillations and
OSARGs. Thus our catalog comprise variables of the same class but with
larger amplitudes and luminosities (Dziembowski and Soszyñski 2010).

\begin{figure}[p]
\centerline{\includegraphics[width=12.2cm, bb=55 55 530 745]{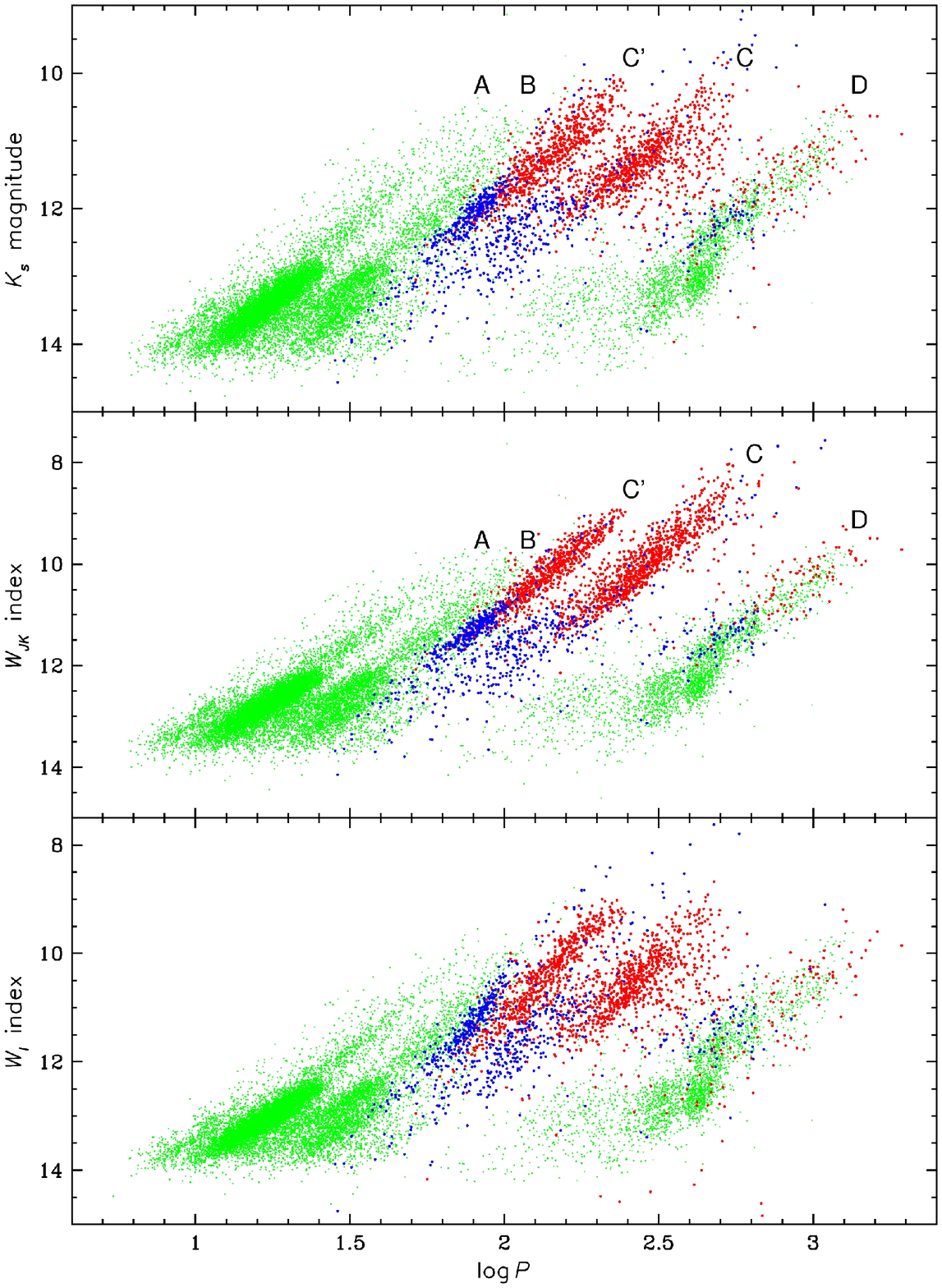}}
\FigCap{Period--luminosity diagrams for LPVs in the SMC. 
{\it Upper, middle} and {\it bottom panels} show $\log{P}$--$K_s$,
$\log{P}$--$W_{JK}$ and $\log{P}$--$W_I$ diagrams, respectively. Green
points mark OSARG variables, blue points indicate O-rich Miras and SRVs,
and red points show positions of C-rich Miras and SRVs. Each star is
represented by one (the primary) period.}
\end{figure}

\Section{Discussion}

Our catalog contains the largest sample of LPVs in the SMC ever
published. It is natural to compare the red giant population in both
Magellanic Clouds using the OGLE-III catalog of LPVs in the LMC
(Paper~I). Fig.~5 shows the relation between logarithm of the period and
$K_s$, $W_{JK}$ and $W_I$ magnitudes. Variable stars are distributed on a
series of different parallel relations. These relations were first
discovered by Wood \etal (1999) from the analysis of LPVs in the LMC. The
PL sequence~C is populated by Miras and SRVs. Note that in the
$\log{P}$--$K_s$ plane some the longest-period stars in sequence~C spread
to low magnitudes, while this is not seen in the reddening free
$\log{P}$--$W_{JK}$ diagram. These faint sources, mainly Mira stars, are
surrounded by thick circumstellar dust shells. Their magnitudes in the {\it
I}-band are close to the detection limit of the OGLE project (20-21~mag).

The distribution of SMC SRVs in the PL diagrams, in particular in the
$\log{P}$--$W_{JK}$ plane (the middle panel of Fig.~5) confirms the
existence of an additional PL ridge between sequences C and C$'$. The
status of this sequence is unclear (additional mode of pulsation, different
population of AGB stars?).

\begin{figure}[htb]
\centerline{\includegraphics[width=12.2cm, bb=55 185 555 750]{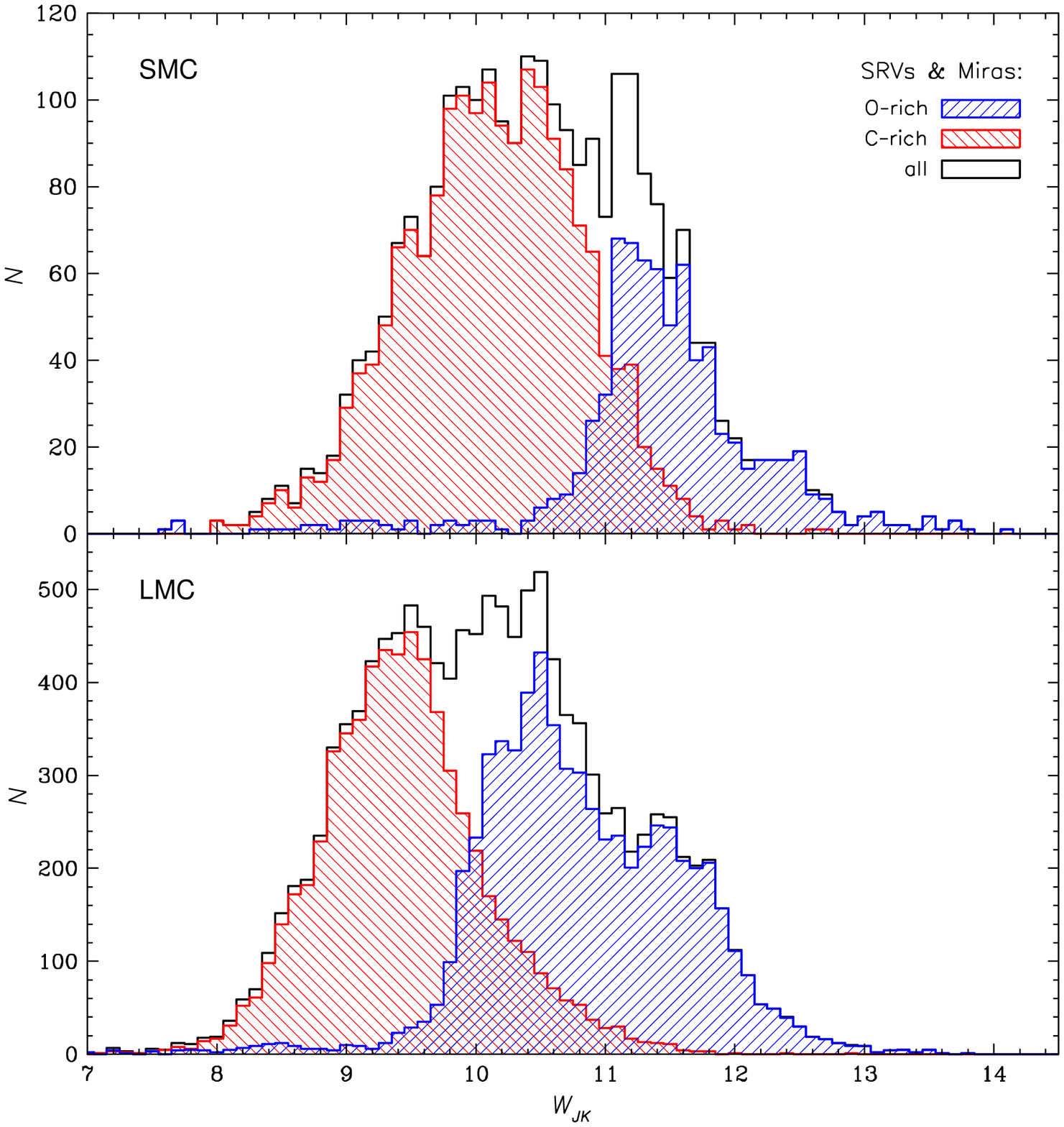}}
\FigCap{Distribution of the reddening-free $W_{JK}$ magnitudes for SRVs and
Miras in the SMC ({\it upper panel}) and in the LMC ({\it lower
panel}). Blue histograms show O-rich giants, red -- C-rich stars and black
-- both spectral classes.}
\end{figure}

The most striking difference between LPVs in both Magellanic Clouds is the
proportion of O-rich and C-rich stars. In Fig.~6 we present the
distribution of $W_{JK}$ magnitudes of SRVs and Miras in both
systems. C-rich stars, which generally dominate among brighter Miras and
SRVs, extend to fainter magnitudes in the SMC than in the LMC. A larger
ratio of C-rich to O-rich stars in the SMC has been known for a long time
(Blanco \etal 1980, Lloyd Evans 1988). This effect is related to the lower
metallicity of the SMC (Iben and Renzini 1983). It is worth noting that the
total distribution of magnitudes for SRVs and Miras are similar in both
galaxies (taking into account different distance moduli). It means that AGB
stars in the SMC become C-rich earlier than in the LMC, but this fact does
not significantly affect the distribution of their reddening-free
near-infrared brightness.

\begin{figure}[htb]
\centerline{\includegraphics[width=12.2cm, bb=55 330 555 750]{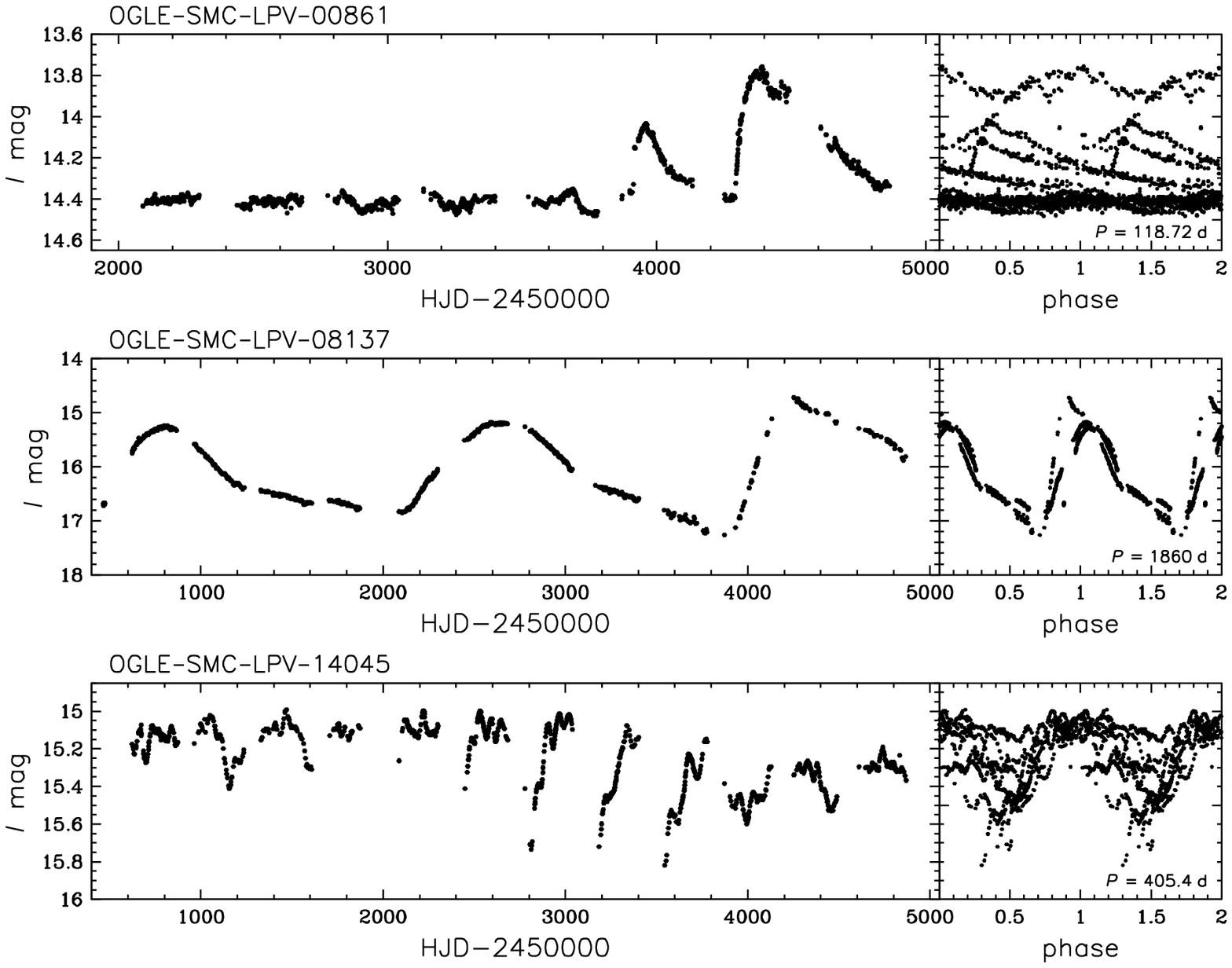}}
\FigCap{Light curves of three interesting objects in our catalog. {\it Left
panels} show unfolded light curves, and {\it right panels} show the same
light curves folded with the primary periods. {\it Upper panels} present a
SRV with outbursts. {\it Middle panels} show the longest-period Mira
OGLE-SMC-LPV-08137. {\it Lower panels} contain a sequence D star with the
changes of the mean luminosity which are correlated with the amplitude
variations.}
\end{figure}

The huge OGLE catalog of LPVs is a source of particularly interesting, very
rarely observed objects and phenomena. The light curves of three such
exceptional stars are shown in Fig.~7. The upper panel presents
OGLE-SMC-LPV-00861 -- a SRV star with two outbursts with amplitudes 0.38
and 0.65 mag. Such a behavior is common in blue variables (Be stars), but
very rare in red giants. It is possible that we observed a process of
swallowing a planet by an expanding red giant (Retter \etal 2006). This
object will be monitored by the OGLE survey in its fourth phase (OGLE-IV).

The middle panel of Fig.~7 shows the longest-period Mira star in our
catalog -- OGLE-SMC-LPV-08137 (IRAS00483-7347) -- with the pulsation period
exceeding 5~years. Actually, this is probably the longest-period Mira known
in any environment. During 14 years of observations by the OGLE survey we
covered only three maxima of this object. Groenewegen \etal (2009)
considered this object as a good candidate for a super-AGB star, \ie an
8-11~$M_\odot$ counterpart of AGB stars.

Finally, the lower panel of Fig.~7 presents the light curve of a sequence~D star, OGLE-SMC-LPV-14045. Our catalog contains at least two thousand OSARGs and SRVs with the primary periods corresponding to sequence~D. Many more stars have secondary periods that appear in this sequence. The nature of these long-period variations is not yet understood, and the large samples of the sequence D stars detected in both Magellanic Clouds may give a clue to solve this puzzle. OGLE-SMC-LPV-14045 is an example of a sequence~D star with a variable amplitude of the long-period variations. When the amplitude exceeded a certain value, the luminosity of this star decreased also in the maximum light. Such a behavior may be expected, if we assume that the sequence~D variations are caused by a cloud of matter orbiting the star. During the episodes of intensified mass loss from the giant, the cloud becomes denser and larger, which increases the amplitude of variations, and finally the whole star is obscured by dust, which decreases its mean luminosity.

\Acknow{We wish to thank W.~Dziembowski for reading the manu\-script and
providing very useful suggestions. We are grateful to Z.~Ko³acz\-kowski,
G. Pojmañski and J.~Skowron for providing software and data which enabled
us to prepare this study.

The research leading to these results has received funding from the
European Research Council under the European Community's Seventh Framework
Program\-me (FP7/2007-2013)/ERC grant agreement no. 246678. This study made use of the {\sc Simbad} database, operated at CDS, Strasbourg, France.}


\begin{references}
\refitem{Alard, C., and Lupton, R.H.}{1998}{\ApJ}{503}{325}
\refitem{Artyukhina, N.M. \etal}{1995}{~}{~}{``General Catalogue of Variable Stars'', 4rd Ed., vol.~V. Extragalactic Variable Stars, ``Kosmosinform'', Moscow}
\refitem{Blanco, V.M., McCarthy, M.F., and Blanco, B.M.}{1980}{\ApJ}{242}{938}
\refitem{Cioni, M.-R.L., Blommaert, J.A.D.L., Groenewegen, M.A.T., Habing, H.J., Hron, J., Kerschbaum,~F., Loup, C., Omont, A., van Loon, J.T., Whitelock, P.A., and Zijlstra, A.A.}{2003}{\AA}{406}{51}
\refitem{Cutri, R.M., \etal}{2003}{~}{~}{``2MASS All-Sky Catalog of Point Sources''}
\refitem{Dartayet, M., and Landi Dessy, J.}{1952}{\ApJ}{115}{279}
\refitem{De Ridder, J., Barban, C., Baudin, F., Carrier, F., Hatzes, A.P., Hekker, S., Kallinger, T., Weiss, W.W., Baglin, A., Auvergne, M., Samadi, R., Barge, P., and Deleuil, M.}{2009}{Nature}{459}{398}
\refitem{Dziembowski, W.A., and Soszy{\'n}ski, I.}{2010}{\AA}{524}{88}
\refitem{Gilliland, R.L., \etal}{2010}{\PASP}{122}{131} 
\refitem{Groenewegen, M.A.T.}{2004}{\AA}{425}{595}
\refitem{Groenewegen, M.A.T., Sloan, G.C., Soszy{\'n}ski, I., and Petersen, E.A.}{2009}{\AA}{506}{1277}
\refitem{Hoffleit, D.}{1935}{Harv. Coll. Obs. Bull.}{900}{3}
\refitem{Iben, I., Jr., and Renzini, A.}{1983}{ARA\&A}{21}{271}
\refitem{Ita, Y., Tanab{\'e}, T., Matsunaga, N., Nakajima, Y., Nagashima, C., Nagayama, T., Kato, D., Kurita, M., Nagata, T., Sato, S., Tamura, M., Nakaya, H., and Nakada, Y.}{2004a}{\MNRAS}{347}{720}
\refitem{Ita, Y., Tanab{\'e}, T., Matsunaga, N., Nakajima, Y., Nagashima, C., Nagayama, T., Kato, D., Kurita, M., Nagata, T., Sato, S., Tamura, M., Nakaya, H., and Nakada, Y.}{2004b}{\MNRAS}{353}{705}
\refitem{Kato, D., \etal}{2007}{PASJ}{59}{615}
\refitem{Kiss, L.L., and Bedding, T.R.}{2004}{\MNRAS}{347}{L83}
\refitem{Lah, P., Kiss, L.L., and Bedding, T.R.}{2005}{\MNRAS}{359}{L42}
\refitem{Lloyd Evans, T.}{1971}{~}{~}{in: ``The Magellanic Clouds'', ed. A.B. Muller (Reidel, Dordrecht), p.~74}
\refitem{Lloyd Evans, T.}{1978a}{\MNRAS}{183}{305}
\refitem{Lloyd Evans, T.}{1978b}{\MNRAS}{183}{319}
\refitem{Lloyd Evans, T., Glass, I.S., and Catchpole, R.M.}{1988}{\MNRAS }{231}{773}
\refitem{McKibben Nail, V.}{1942}{Ann. Harv. Coll. Obs.}{109}{27}
\refitem{Payne-Gaposchkin, C., and Gaposchkin, S.}{1966}{Smithsonian Contrib. Astrophys.}{9}{1}
\refitem{Pojmañski, G.}{1997}{\Acta}{47}{467}
\refitem{Raimondo, G., Cioni, M.-R.L., Rejkuba, M., and Silva, D.R.}{2005}{\AA}{438}{521}
\refitem{Retter, A., Zhang, B., Siess, L., and Levinson, A.}{2006}{\MNRAS}{370}{1573} 
\refitem{Schechter, P.L., Mateo, M., and Saha, A.}{1993}{\PASP}{105}{1342}
\refitem{Schultheis, M., Glass, I.S., and Cioni, M.-R.}{2004}{\AA}{427}{945}
\refitem{Shapley, H., Yamamoto, I., and Wilson, H.H.}{1925}{Harv. Coll. Obs. Circ.}{280}{1}
\refitem{Shapley, H., and McKibben Nail, V.}{1951}{Proc. Nat. Acad. Sci.}{37}{138}
\refitem{Soszyñski, I., Udalski, A., Kubiak, M., Szymañski, M., Pietrzyñski, G., \.Zebruñ, K., Szewczyk,~O., and Wyrzykowski, {\L}.}{2004}{\Acta}{54}{129}
\refitem{Soszyñski, I., Udalski, A., Kubiak, M., Szymañski, M., Pietrzyñski, G., \.Zebruñ, K., Szewczyk,~O., Wyrzykowski, {\L}., and Ulaczyk, K.}{2005}{\Acta}{55}{331}
\refitem{Soszyñski, I., Dziembowski, W.A., Udalski, A., Kubiak, M., Szymañski, M.K., Pietrzyñski, G., Wyrzykowski, {\L}., Szewczyk,~O., and Ulaczyk, K.}{2007}{\Acta}{57}{201}
\refitem{Soszyñski,~I., Udalski, A., Szymañski,~M.K., Kubiak,~M., Pietrzyñski,~G., Wyrzykowski,~£., Szewczyk,~O., Ulaczyk,~K., and Poleski,~R.}{2009}{\Acta}{59}{239 (Paper~I)}
\refitem{Szymañski, M.K.}{2005}{\Acta}{55}{43}
\refitem{Tabur, V., Bedding, T.R., Kiss, L.L., Giles, T., Derekas, A., and Moon, T.T.}{2010}{\MNRAS}{409}{777}
\refitem{Thackeray, A.D.}{1958}{\MNRAS}{118}{117}
\refitem{Udalski, A.}{2003}{\Acta}{53}{291}
\refitem{Udalski, A., Szymañski, M.K., Soszyñski, I., and Poleski, R.}{2008}{\Acta}{58}{69}
\refitem{Wood, P.R., \etal (MACHO team)}{1999}{~}{~}{in: {\it IAU Symp.} 191, ``Asymptotic Giant Branch Stars'', Ed. T. Le~Bertre, A. L\'ebre, and C. Waelkens (San Francisco: ASP), p.~151}
\refitem{Wo¼niak, P.R.}{2000}{\Acta}{50}{421}
\refitem{Wray, J.J., Eyer, L., and Paczyñski, B.}{2004}{\MNRAS}{349}{1059}
\end{references}
\end{document}